\documentclass{article}
\usepackage{spconf,amsmath,graphicx}
\usepackage[hang,flushmargin]{footmisc} 
\usepackage{nicefrac}

\usepackage{amssymb,epsfig, url}
\usepackage{bbm}
\usepackage{enumitem}
\usepackage[noadjust]{cite}
\usepackage{makecell}
\usepackage{tikz}
\usepackage{changepage}   

\newcommand {\myvec}[1] {{\mbox{\boldmath $#1$}}}

\newcommand {\us} {\myvec{s}}
\newcommand {\ub} {\myvec{b}}
\newcommand {\uy} {\myvec{y}}
\newcommand {\Cset} {\mathbb{C}}
\newcommand {\Rset} {\mathbb{R}}

\newcommand {\Nset} {\mathbb{N}}

\newcommand{\optarg}[1][]{%
  \ifthenelse{\isempty{#1}}%
    {}
    {(((#1)))}
}

\newcommand {\Ksc} {K_{\rm{sc}}}
\newcommand {\Tcp} {T_{\rm{cp}}}

\definecolor{crimson} {RGB}{220,20,62}
\newcommand*\circled[1]{\tikz[baseline=(char.base)]{
		\node[circle,draw,color=blue, opacity=0.9,inner sep=1pt] (char) {\footnotesize #1};}}

\setenumerate[1]{label=\arabic*}
\setenumerate[2]{label*=.\arabic*}
\setenumerate[3]{label*=.\arabic*}
\setenumerate{align=right}
\setenumerate{leftmargin =*}

\title{On Neural Architectures for Deep Learning-based Source Separation of Co-Channel OFDM Signals}
\name{Gary C.F. Lee \quad 
    Amir Weiss \quad 
    Alejandro Lancho  \quad 
    Yury Polyanskiy  \quad 
    Gregory W. Wornell}
\address{Massachusetts Institute of Technology}

\begin{document}

\maketitle

\begin{abstract}
We study the single-channel source separation problem involving orthogonal frequency-division multiplexing (OFDM) signals, which are ubiquitous in many modern-day digital communication systems. Related efforts have been pursued in monaural source separation, where state-of-the-art neural architectures have been adopted to train an end-to-end separator for audio signals (as 1-dimensional time series). In this work, through a prototype problem based on the OFDM source model, we assess---and question---the efficacy of using audio-oriented neural architectures in separating signals based on features pertinent to communication waveforms. Perhaps surprisingly, we demonstrate that in some configurations, where \emph{perfect} separation is theoretically attainable, these audio-oriented neural architectures perform poorly in separating co-channel OFDM waveforms. Yet, we propose critical domain-informed modifications to the network parameterization, based on insights from OFDM structures, that can confer about $30$ dB improvement in performance. 
\end{abstract}
\vspace{-0.03cm}
{\let\thefootnote\relax\footnotetext{\scriptsize 
Research was sponsored by the United States Air Force Research Laboratory and the United States Air Force Artificial Intelligence Accelerator and was accomplished under Cooperative Agreement Number FA8750-19-2-1000. The views and conclusions contained in this document are those of the authors and should not be interpreted as representing the official policies, either expressed or implied, of the United States Air Force or the U.S. Government. The U.S. Government is authorized to reproduce and distribute reprints for Government purposes notwithstanding any copyright notation herein.\\
This work is also supported by the NSF under Grant No. CCF-2131115.\\
Alejandro Lancho has received funding from the EU’s Horizon 2020 research and innovation programme under the Marie Sklodowska-Curie grant agreement No. 101024432. \\
The authors acknowledge the MIT SuperCloud and Lincoln Laboratory Supercomputing Center for providing HPC resources that have contributed to the research results reported within this paper.
}}
\vspace{-0.3cm}
\section{Introduction}
\label{sec:intro}
\vspace{-0.3cm}

Source separation is a long-standing problem of interest in many engineering applications. Particularly challenging is the \emph{single-channel} source separation (SCSS) setting that has recently gained more interest \cite{benaroya2005audio, grais2014deep, qian2018past, blouet2008evaluation}, corresponding to an underdetermined scenario, in the absence of spatial diversity. Most prominently, SCSS with audio signals has received much attention \cite{wang2018supervised, purwins2019deep, huang2014deep}. State-of-the-art solutions benefit from deep learning approaches, many of which propose novel neural architectures to achieve improved separation ability. For example, the use of convolutional \cite{stoller2018wave,luo2019conv, tzinis2020sudo}, recurrent \cite{luo2020dual} and attention-based \cite{chen2020dual} layer structures in the separation neural network (NN) have been proposed, with varying degree of success. 
Implicit to these methods are strategies to exploit properties of typical audio signals. In fact, it is believed that the features exploited by state-of-the-art neural architectures are related to separability in the time-frequency space \cite{li2022sound}.

Beyond audio applications, source separation finds important relevance to other domains, such as in radio frequency (RF) and optical systems, for communication and sensing applications \cite{rfchallenge, lancho2022data, sharma2006signal, zhao2021single, tao2019mitigation, ma2020blind, li2014channel}. 
Across these domains, the raw data are 1-dimensional time series. 
Yet, the properties of RF/optical waveforms differ from audio signals---e.g., these signals tend to pack a large amount of information into a finite frequency band (or channel), rendering them no longer sparse in the time-frequency space. In fact, they may be overlapping in this space---described as ``co-channel''. While audio-oriented NNs can work on time series inputs, and could also be just as successful with other modalities, e.g., demonstrated with seismic signals in \cite{novoselov2022sedenoss}), it is uncertain if the same neural architectures are also effective at separating communication signals.

Of particular interest to our work is one such type of RF/optical signals, which modulates information through orthogonal frequency-division multiplexing (OFDM).
While OFDM waveforms form a subclass of digitally modulated signals, it is one of the most ubiquitous modulation found in modern data communications, for both wireless systems (WiFi and 4G/LTE/5G) and optical systems.
We also empirically observe these waveforms to be a challenging class of signals to tackle in data-driven SCSS \cite{lancho2022data, lee2022exploiting}. 

This work focuses on the SCSS of OFDM signals, and the relevant neural architectural choices to capture informative features for signal separation. 
In particular, we consider a prototype problem based on the OFDM model, posed such that perfect separation of the signals is technically attainable through the fast Fourier transform (FFT) algorithm with appropriately chosen parameters. Yet, \emph{without prior knowledge} about the signal model, achieving this is not a simple task. Under this setup, we study whether NN-based approaches can learn to exploit the underlying OFDM structures for SCSS.

To the best of our knowledge, this work is the first to assess the performance (and therefore, the ineffectiveness in some regimes) of neural architectures from audio separation when applied to OFDM waveforms, serving as an important benchmark. We also propose modifications, inspired by OFDM structures, that confer orders of magnitude improvement to the separation performance. The key takeaways are to demonstrate how distinct and challenging digital communication signals can be for existing neural methods in SCSS, and to propose judicious adaptations in advancing neural methods for time-domain signals beyond the efforts in audio domain.

\vspace{-0.3cm}
\section{Problem Formulation}
\label{sec:problem}
\vspace{-0.3cm}

Consider an observed 1-dimensional mixture of signals
{\setlength{\belowdisplayskip}{4pt} \setlength{\belowdisplayshortskip}{4pt}
\setlength{\abovedisplayskip}{2.5pt} \setlength{\abovedisplayshortskip}{2.5pt}
\begin{equation}
\label{eq:dftsep}
    \uy = \us + \ub,
\end{equation}
}where $\us\triangleq[s[0] \, \hdots \, s[N-1]]^{{\rm T}}\in\Cset^N$ is our signal-of-interest (SOI) to be extracted, and $\ub\triangleq[b[0] \, \hdots \, b[N-1]]^{{\rm T}}\in\Cset^N$ is the interference (signal-not-of-interest). The goal is to separate $\us$ from $\ub$, namely, estimate $\us$ from $\uy$ with minimum mean squared error (MSE) as the criterion. 
We assume that the models for $\us$ and $\ub$ are not known; however, we have a dataset of $M$ independent, identically distributed examples, $\{(\uy^{(i)},\us^{(i)})\}_{i=1}^M$. This setup naturally lends itself to a data-driven approach for the SCSS problem. 

In this work, we consider an SOI and an interference that are discrete-time OFDM waveforms, formally expressed as
{\setlength{\belowdisplayskip}{5pt} 
\setlength{\belowdisplayshortskip}{5pt}
\setlength{\abovedisplayskip}{5pt} 
\setlength{\abovedisplayshortskip}{5pt}
\begin{equation}
\begin{aligned}
\label{eq:ofdmsources}
    &s[n] = \sum_{p=0}^{P-1}\sum_{k=0}^{K-1} g_{k,p} \, r[n-p\cdot(K+\Tcp)-\Tcp, \, k], \\[-0.2em]
    &b[n] = \sum_{p=0}^{P-1}\sum_{k=0}^{K-1} h_{k,p} \, r[n-p\cdot(K+\Tcp)-\Tcp, \, k], \\[-0.2em]
    &r[n, \, k] \triangleq \exp({\nicefrac{j 2\pi kn}{K}}) \,  \mathbbm{1}_{\left\{-\Tcp \leq n < K\right\}},
\end{aligned}
\end{equation}
}for $n\in\{0,\ldots,N-1 \}$, where $K\in2\cdot\Nset$ is the total number of orthogonal complex sinusoid terms (also termed as subcarriers; this also corresponds to the FFT size). The coefficients $g_{k,p}\in\mathcal{G}$, $h_{k,p}\in\mathcal{H}$ are the modulated symbols, and $\mathcal{G}, \mathcal{H}$ are their alphabets (constellations), respectively. A cyclic prefix (CP) is typically added before an OFDM symbol. Hence, each OFDM symbol is described for the interval $[-\Tcp, K]$, where $\Tcp\in\Nset$ is the CP length, and $K$ is the OFDM symbol length (without CP). The signals span $P$ OFDM symbols, and their individual finite support is reflected by the finitely supported function $r[n,\,k]$. 

In this setting, the observed mixture can also be viewed as an OFDM waveform, with the coefficients being elements from the superconstellation of the SOI's and interference's symbols, i.e., the Minkowski sum $\mathcal{A}\triangleq\mathcal{G}\oplus \mathcal{H}$, such that
{\setlength{\belowdisplayskip}{4pt} 
\setlength{\belowdisplayshortskip}{4pt}
\setlength{\abovedisplayskip}{4pt} 
\setlength{\abovedisplayshortskip}{4pt}
\begin{equation}
\begin{aligned}
\label{eq:ofdmobserved}
    y[n] &= \sum_{p=0}^{P-1}\sum_{k=0}^{K-1} a_{k,p} \, r[n-p\cdot(K+\Tcp)-\Tcp, \,k], \\[-0.2em]
    a_{k,p} &= g_{k,p} + h_{k, p}, \quad a_{k,p}\in\mathcal{A}.
\end{aligned}
\end{equation}
}The existence of a surjective function $f: \mathcal{A}\rightarrow\mathcal{G}$, i.e., every element in $\mathcal{A}$ can be uniquely associated with an element in the SOI's constellation $\mathcal{G}$, suffices for perfect separability.\footnote{Alternatively, a surjective function $f: \mathcal{A}\rightarrow\mathcal{H}$.}

\subsection{Special Case: Real-valued OFDM Signals}\label{subsec:specialcase}
\vspace{-0.25cm}
In order to focus on the core aspects of this problem, namely the underlying Fourier structures and finite coefficient sets present in OFDM, we propose to examine the following simplified (yet not simplistic) special case. Consider \eqref{eq:ofdmsources} with $P=1$,  $\Tcp=N-K$ and $N\in K\cdot\Nset$, namely,
{\setlength{\belowdisplayskip}{2pt} 
\setlength{\belowdisplayshortskip}{2pt}
\setlength{\abovedisplayskip}{2pt} 
\setlength{\abovedisplayshortskip}{2pt}
\begin{equation}
\label{eq:fouriersources}
    s[n] = \sum_{k=0}^{K-1} \underbrace{g_{k,0}}_{\triangleq g_k} r[n-\Tcp,k]= \sum_{k=0}^{K-1} g_{k} \exp({\nicefrac{j 2\pi kn}{K}}), 
\end{equation}
}and similarly for $b[n]$, $y[n]$ with coefficients $h_k$, $a_k$, respectively, such that the (periodic extensions of the) SOI and interference are discrete Fourier series.
Further, we impose the conjugate symmetry constraint on the coefficients $g_{k}$, $h_{k}$,
{\setlength{\belowdisplayskip}{2.5pt} 
\setlength{\belowdisplayshortskip}{2.5pt}
\setlength{\abovedisplayskip}{3pt} 
\setlength{\abovedisplayshortskip}{3pt}
\begin{align*}
    \circled{1}\;g_{0} = g_{K/2} = 0 \quad 
    \circled{2}\;g_{k} = g^*_{K-k}, \, \forall k\in\{1,\ldots,\tfrac{K}{2}-1\}, 
\end{align*}
}and similarly for $h_{k}$, where $z^*$ denotes the complex conjugate of a complex number $z$. 
Consequently, the waveforms generated by \eqref{eq:fouriersources} are real-valued: $\us, \ub \in \Rset^N \Rightarrow \uy \in \Rset^N$. 

The purpose of focusing on this special case is to test the ability of candidate neural architectures to capture, or learn to exploit, the subcarriers' \emph{orthogonality} and the \emph{discrete} constellation set, with which perfect separation is attainable.

\begin{table*}[t]
    \caption{MSE (in decibels, dB) of the extracted SOI. Entries with MSE$<10^{-2}$ (i.e., $-20$ dB\protect\footnotemark[4]) are in \textcolor{crimson}{red}; improvement of our proposed architecture, compared to the best-performing benchmark method, is reported in parenthesis in the last row.}
    \centering
    \begin{tabular}{|l|>{\centering}p{0.165\textwidth}| >{\centering}p{0.165\textwidth}| >{\centering}p{0.188\textwidth}| >{\centering\arraybackslash}p{0.165\textwidth}|}
        \hline
         & Case 1 & Case 2 & Case 3 & Case 4  \\[-0.3em]
         & Disjoint & BPSK+BPSK & BPSK+4-PAM (Mixed) & 4-PAM+4-PAM  \\
        \hline
        \hline
         Wave-U-Net \cite{stoller2018wave} & \textcolor{crimson}{$-57.246$ dB} & \textcolor{crimson}{$-46.827$ dB} & $-4.663$ dB & $-4.665$ dB  \\
         \hline
         Conv-TasNet \cite{luo2019conv} & \textcolor{crimson}{$-40.790$ dB} & $-12.179$ dB & $-1.060$ dB & $-1.009$ dB\\
        \hline
         Sudo-Rm-Rf \cite{tzinis2020sudo} & \textcolor{crimson}{$-37.023$ dB} & \textcolor{crimson}{$-26.493$ dB} & $-12.855$ dB & $-11.495$ dB\\
         \hline
         Dual Path RNN \cite{luo2020dual} & \textcolor{crimson}{$-41.425$ dB} & \textcolor{crimson}{$-27.302$ dB} & $-0.671$ dB & $-0.542$ dB \\
         \hline
         DPTNet \cite{chen2020dual} & \textcolor{crimson}{$-36.825$ dB} & \textcolor{crimson}{$-33.652$ dB} & $-3.548$ dB & $-2.432$ dB\\
         \hline
         \hline
         \textbf{Modified Wave-U-Net} & \textcolor{crimson}{\textbf{$\mathbf{-65.526}$ dB}} & \textcolor{crimson}{\textbf{$\mathbf{-47.558}$ dB}} & \textcolor{crimson}{\textbf{$\mathbf{-47.377}$ dB}} & \textcolor{crimson}{\textbf{$\mathbf{-41.156}$ dB}}\\[-0.2em]
         {\small
         \textbf{(Proposed)}} &  {\small($\downarrow 8.280$)} & 
         {\small($\downarrow 0.731$)} & 
         {\small($\downarrow 34.522$)} & 
         {\small($\downarrow 29.661$) }\\
         \hline
    \end{tabular}
    \vspace{-0.6cm}
    \label{tab:sinusoidsepdnn}
\end{table*}

\vspace{-0.4cm}
\section{Regimes of Methodology}
\label{sec:method}
\vspace{-0.33cm}

\begin{figure}
    \centering
    \begin{tabular}{cc}
    \includegraphics[width=0.21\textwidth, trim={0.6cm 0 0.25cm 0}]{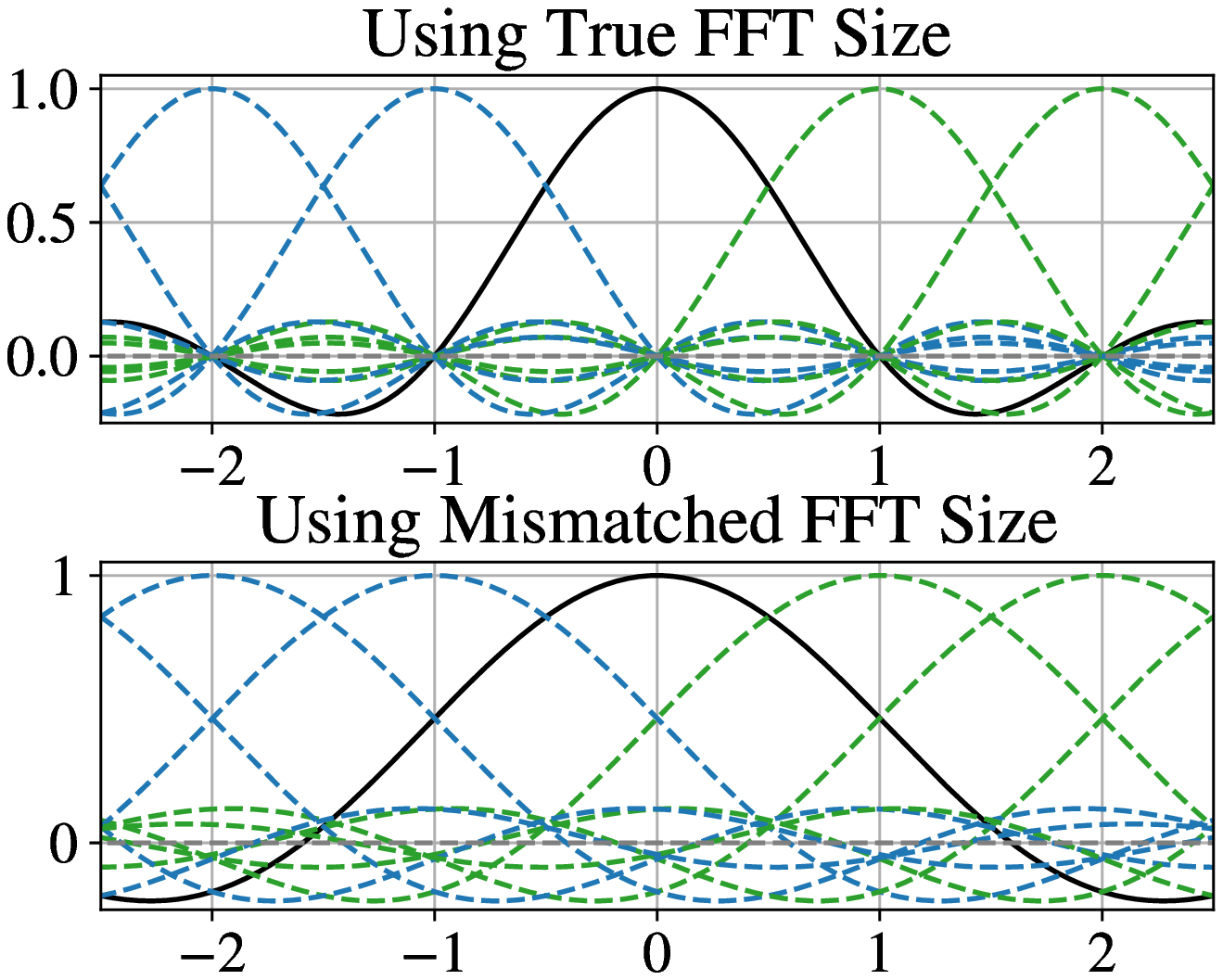} &
    \includegraphics[width=0.22\textwidth, trim={0.25cm 0 0.5cm 0}]{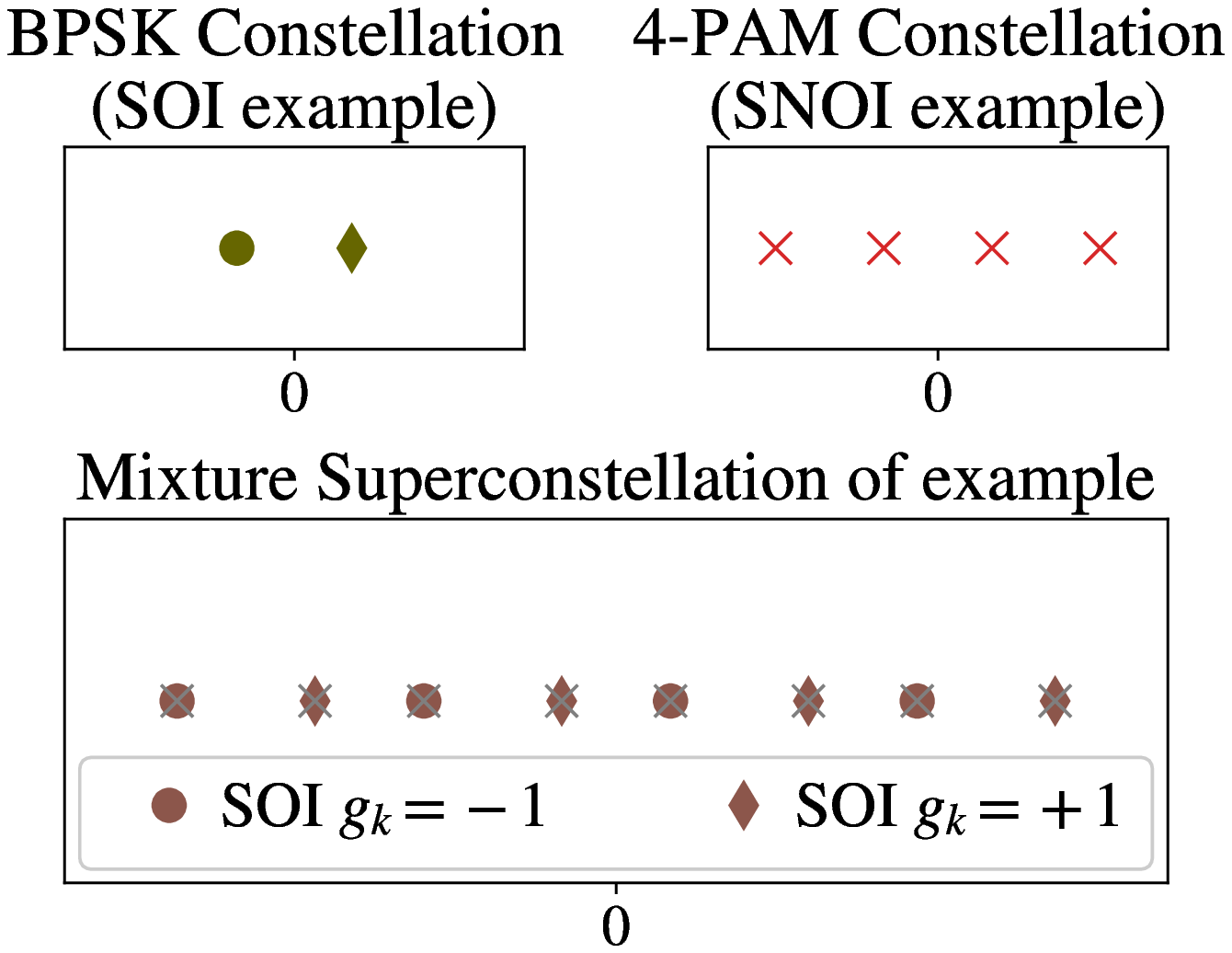} \\[-0.4em]
    (i) & (ii) 
    \end{tabular}
    \vspace{-0.35cm}
    \caption{Visualization of OFDM structure---(i) using the appropriate FFT size leads to orthogonality between subcarriers; a mismatched FFT leads to a loss of orthogonality at the subcarrier frequencies; (ii) for an appropriate choice of discrete constellations, a surjective mapping of points from the superconstellation to an SOI symbol can be obtained.}
    \label{fig:ofdmviz}\vspace{-0.5cm}
\end{figure}

Before we consider data-driven methods and neural architectures for separating OFDM signals of the form \eqref{eq:fouriersources}, we briefly discuss conventional (informed) model-based approaches for reference. Consider a frequency domain approach (i.e., using FFT), where the SOI is reconstructed from the mixture's spectral coefficients, $\{a_k\}_{k=0}^{K-1}$. The first step requires an \emph{exact} FFT size, which can be unknown, that preserves the orthogonality of the subcarriers. Otherwise, the spectral ``leakage" from neighboring subcarriers will typically result in a significantly larger superconstellation, due to the loss of orthogonality (illustrated in Fig.~\ref{fig:ofdmviz}(i)). Of course, for this, one has to establish a realizable method that extracts the underlying spectral coefficients---corresponding to the surjective mapping of the mixture's symbol superconstellation to the SOI's constellation points (e.g. Fig.~\ref{fig:ofdmviz}(ii)). The existence of such a routine, albeit through explicit knowledge of source model parameters, demonstrates a possible approach to achieve perfect signal separation performance for \eqref{eq:fouriersources}. However, this does not correspond to a practicable algorithm in a more general sense, e.g., with partial knowledge about the source models. 

On the other hand, monaural separation architectures can be used for an end-to-end signal separator \emph{without} explicitly requiring the source models. Since the latent sources in the special case considered in Section~\ref{subsec:specialcase} are real-valued, we can adopt the neural architectures proposed in audio source separation works, developed for real-valued time domain inputs.

For these audio-oriented methods, information pertaining to the sources being a discrete Fourier series (or an OFDM waveform) is not explicitly utilized.
Clearly, an effective architecture for this SCSS problem ought to learn to exploit properties relating to the OFDM's subcarrier structure and discrete symbol constellations from data. Intriguingly, we demonstrate that beyond a limited regime of this problem, audio-based neural architectures fail to separate OFDM waveforms that are in fact perfectly separable. We then propose certain domain-informed modifications that revive these architectures, consequently leading to successful separation---improving the figure of merit (MSE) by orders of magnitudes.

For the rest of this work, we focus on computational experiments with parameters $N=4096$, $K=64$, $\Ksc=28$, where $\Ksc$ corresponds to the (maximum) number of unique nonzero coefficients (subcarriers) in this model.\footnote{These parameters are based on 802.11n WiFi waveform properties \cite{ieee80211n}.} We consider $4$ different cases of $g_k$, $h_k$, for $k\in\{1,\ldots,\Ksc\}$:
\\[-1em]
\begin{adjustwidth}{0.25cm}{}
\textbf{Case 1: Disjoint frequency sets:} $g_k=0$ when $h_k\neq0$ and \emph{vice versa}, where nonzero indices are randomly chosen once, and stay fixed thereafter. The nonzero coefficients are drawn from a random continuous uniform distribution, $g_k\sim \mathcal{U}[-\sqrt{3},\sqrt{3}]$, $h_k\sim \mathcal{U}[-4\sqrt{3},4\sqrt{3}]$. \\[0.35em] 
\textbf{Case 2: ``BPSK\footnote{\label{abbr}BPSK: Binary Phase Shift Keying; 4-PAM: 4 Pulse-Amplitude Modulation; these are modulation schemes typical in digital communication signals.}-like'' coefficients:} $g_k\in\{+1, -1\}$ and $h_k\in\{+4, -4\}$. \\[0.35em]
\textbf{Case 3: ``Mixed'' coefficients:} $g_k\in\{+1, -1\}$ and $h_k\in\{\nicefrac{+12}{\sqrt{5}},$ $\nicefrac{+4}{\sqrt{5}}, \nicefrac{-4}{\sqrt{5}}, \nicefrac{-12}{\sqrt{5}}\}$. \\ [0.35em]
\textbf{Case 4: ``4-PAM\footnotemark[3]-like'' coefficients:} $g_k\in\{\nicefrac{+3}{\sqrt{5}}, \nicefrac{+1}{\sqrt{5}},$ $ \nicefrac{-1}{\sqrt{5}}, \nicefrac{-3}{\sqrt{5}}\}$ and $h_k\in\{\nicefrac{+12}{\sqrt{5}}, \nicefrac{+4}{\sqrt{5}}, \nicefrac{-4}{\sqrt{5}}, \nicefrac{-12}{\sqrt{5}}\}$.\\[-1.1em]
\end{adjustwidth}
We introduce the appropriate scaling factors on the source components such that the SOI $\us$ has unit average power.
We set $g_0=h_0=0$, and $g_k=h_k=0$ for $k\in[\Ksc+1, K/2]$, and recall that for $k>K/2$, the coefficients are constrained to have conjugate symmetry. 
In all cases, the average interference power is 16 times that of the average SOI power.

\footnotetext[4]{\label{msethr}An approximation of the best separation performance reported in \cite{stoller2018wave, tzinis2020sudo, luo2019conv, luo2020dual, chen2020dual}.}
\setcounter{footnote}{4}

\vspace{-0.35cm}
\section{Neural Architectures for SCSS}
\label{sec:experiment}
\vspace{-0.3cm}

We now describe the details of our experiments. We then present the separation performance, and compare how they fare in the 4 different cases established earlier.

\vspace{-0.4cm}
\subsection{Implementation Details}
\vspace{-0.25cm}
To the best of our knowledge, there are no established baseline NN methods for this OFDM SCSS problem. Hence, part of our work is to train selected state-of-the-art NNs from audio separation \cite{stoller2018wave, tzinis2020sudo, luo2019conv, luo2020dual, chen2020dual} for this problem, and assess their performance to serve as our comparison benchmark. Asteroid, the PyTorch-based audio source separation toolbox \cite{pariente2020asteroid}, is used for state-of-the-art neural architectures, whereas Wave-U-Net and its modified version (our proposed architecture, detailed later) are implemented in PyTorch.\footnote{Repository containing code and implementation details: \url{https://github.com/RFChallenge/SCSS_OFDMArchitecture}.} The training and validation sets comprise $90,000$ and $10,000$ independent realizations of mixture-SOI pairs respectively. Adam optimizer with a learning rate $10^{-4}$ is used to train the respective NNs for $2\times10^3$ epochs, with early stopping after $100$ epochs of no improvement on validation. 
In Table~\ref{tab:sinusoidsepdnn} we report the MSE performance of the selected neural architectures in the reconstruction of the SOI, on a test set comprising $10^3$ examples.

\vspace{-0.4cm}
\subsection{Performance of Audio-Oriented NNs}
\vspace{-0.25cm}
As expected, the audio-oriented NN models are all effective in separating signals with disjoint frequency sets (Case 1), i.e., masking/filtering the frequencies that make up the SOI $\us$. Yet, we see that many of these neural architectures can still separate highly co-channel signals under some cases, as in Case 2 with OFDM waveforms bearing BPSK symbols. The models' success in Case 2 indicates that the mechanisms of these audio-based neural methods have the potential to separate signals that overlap in time and frequency, i.e., deviating from the proximity characteristics indicated in \cite{li2022sound}, thereby reflecting the potential generalizability of some of these architectures. Nonetheless, as we advance to consider more complex constellation, such as 4-PAM (Cases 3 and 4), these architectures are no longer as effective at separating these signals when trained with the configuration considered here. 

\begin{table}[t]
    \vspace{-0.25cm}
    \caption{MSE (in dB) of the extracted SOI using the modified Wave-U-Net with different first-layer kernel sizes.}
    \centering
    \begin{tabular}{|l|r||l|r|}
        \hline
        Kernel Size & MSE & Kernel Size & MSE\\
        \hline
        \hline
        $W=15$ & $-6.030$ dB & $W=65$ & $-42.824$ dB\\
        \hline
        $W=21$ & $-5.621$ dB & $W=71$ & $-42.099$ dB \\
        \hline
        $W=31$ & $-6.183$ dB & $W=81$ & $-42.690$ dB \\
        \hline
        $W=51$ & $-16.319$ dB & $W=101$ & $-41.156$ dB \\ 
        \hline
        $W=63$ & $-41.380$ dB & $W=201$ & $-44.319$ dB \\
        \hline
    \end{tabular}
    \vspace{-0.66cm}
    \label{tab:kernelsize}
\end{table}

\vspace{-0.3cm}
\section{OFDM Domain-informed Architecture}
\label{sec:ofdmmod}
\vspace{-0.2cm}
We now propose modifications to one of the architectures based on insights from OFDM signals.
Thereafter, we review possible justifications for the improvement attained by drawing connections to OFDM's Fourier structures, which in turn leads to guidelines for domain-informed parameterization. 

\vspace{-0.4cm}
\subsection{Proposed Neural Architecture Modifications}
\vspace{-0.2cm}
Referring to the model-based approach, we seek an NN that is capable of approximating an appropriately sized FFT operator (Fig.~\ref{fig:ofdmviz}). Based on this insight, a natural modification to the NN is to increase the number of filters ($20\times$ as many) and the receptive fields of these filters on the first layer (kernel size $W=101$), which operates on the time-domain input. We introduce these modifications to Wave-U-Net---the simplest among those investigated. The last row of Table~\ref{tab:sinusoidsepdnn} reports the substantial improvement in MSE due to these modifications.

To further lend credence to the role of first-layer kernel size, we show the SCSS results on Case 4 using the modified Wave-U-Net with different sizes in Table~\ref{tab:kernelsize}. Here, we see a significant improvement in separation performance when kernel sizes $63$ and longer are used in the modified Wave-U-Net  (recalling that the true FFT size $K=64$). 

\vspace{-0.4cm}
\subsection{Discussion on Fourier Structures}
\label{sec:structures}
\vspace{-0.2cm}

Referring to \eqref{eq:fouriersources}, each time-domain sample is a sum of statistically independent random variables; by the central limit theorem (CLT), each of these samples is marginally Gaussian distributed as $K\to\infty$. Note that even when a (fixed) $W\ll K$ time samples are considered, they are asymptotically jointly Gaussian, unless they are exactly a period away \cite{schlezinger2014fresh}. Yet, $K$ consecutive time-domain samples are not jointly Gaussian, as evident from the discrete (non-Gaussian) coefficient set.\footnote{Indeed, the CLT cannot be invoked in this case.} A large receptive field in the NN, particularly on the raw input, may be related to the window (FFT) size where one is able to capture the underlying non-Gaussianity. 

\begin{figure}
    \centering
    \includegraphics[width=0.44\textwidth, trim={0 0.05cm 0 0.4cm}]{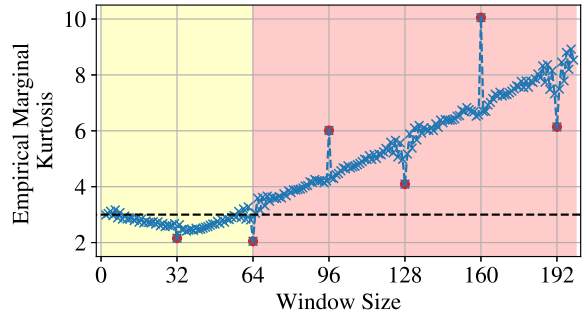}
    \vspace{-0.5cm}
    \caption{Empirical marginal kurtosis of the latent representation (real part of FFT coefficients) versus window length $W$.}
    \label{fig:kernelkurtosis}
    \vspace{-0.4cm}
\end{figure}

To further illustrate this, we consider the empirical marginal kurtosis of the latent representation obtained by taking the $W$-order FFT of some length $W$.\footnotemark[7] Fig.~\ref{fig:kernelkurtosis} shows the empirical kurtosis of the (real part of) coefficients extracted across $2\times10^6$ independent realizations of $\uy$. We highlight two key observations: 1) spikes appear at multiples of $K/2$; and 2) the empirical kurtosis stays close to $3$ when $W<K$ (resembling that of Gaussian random variables), but increases for longer windows, departing from Gaussianity. We note that the latter is true for \eqref{eq:fouriersources}, where a growing window leads to more low-magnitude coefficients (with the same number of non-zero subcarrier symbols), which in turn leads to a heavy-tailed distribution of these latent coefficients. 

\footnotetext[7]{\label{kurtosisica}This is loosely based on kurtosis as a surrogate measure of non-Gaussianity, as typical in algorithms, e.g., independent component analysis (ICA) \cite{comon2010handbook}.}

These observations lead to two possible mechanisms through which a long first-layer kernel is exploiting. First, a long kernel has the representation power to approximate the FFT of the exact order or its harmonics (the red circle data points in Fig.~\ref{fig:kernelkurtosis}). Second, a significantly long kernel can process the signal in a regime where it is non-Gaussian, but not necessarily exactly at the FFT size (the two colored regions in Fig.~\ref{fig:kernelkurtosis}). While the neural separation mechanism and feature explainability are not the focus of this work, these are potentially impactful aspects calling for further investigation.

We recognize that a deep NN ought to have a large effective receptive field through its stacked layers, even if the kernel sizes of individual convolutional layers are short \cite{luo2016understanding}. Yet, we have observed that none of the deep NN models considered are as effective in Case 4, in contrast to what is achieved through significantly long kernel on the first layer, operating \emph{directly} on the input itself. The disparity in performance remains a question for further consideration.

\vspace{-0.3cm}
\section{Concluding Remarks}
\vspace{-0.3cm}
Our work reveals key insights into strategies for neural architecture choices for potential novel OFDM-based systems---specifically, parameterization (mildly) informed by the FFT size, corresponding to sufficiently long kernel sizes on the first convolutional layer. 
Next steps include revisiting the general model \eqref{eq:ofdmsources}, which has more latent parameters that an appropriate NN has to capture for SCSS. Of further interest is studying the mechanism behind the modified Wave-U-Net architecture proposed, and how that translates broadly to a more effective NN architecture for OFDM waveforms.

\bibliographystyle{IEEEbib}

\small{\bibliography{main}}

\end{document}